\documentclass[conference]{IEEEtran}

  	\usepackage[pdftex]{graphicx}
  	\graphicspath{{../pdf/}{../jpeg/}}
	\DeclareGraphicsExtensions{.pdf,.jpeg,.png}

    \usepackage{color}
    \usepackage{soul}
	\usepackage[cmex10]{amsmath}
	\usepackage{mathabx}
        \usepackage{soul}
	\usepackage{algorithmic}
	\usepackage{array}
	\usepackage{mdwmath}
	\usepackage{mdwtab}
	\usepackage{eqparbox}
	\usepackage{url}
\begin{document}

\title{
Actor-Critic Network 
for O-RAN Resource Allocation: xApp Design, Deployment, and Analysis}


 \author{\authorblockN{Mohammadreza Kouchaki and Vuk Marojevic}
 \authorblockA{Electrical and Computer Engineering, Mississippi State University, USA\\mk1682@msstate.edu, vm602@msstate.edu}}

\maketitle

\begin{abstract}
\textcolor{black}{Open Radio Access Network (O-RAN) has introduced an emerging RAN architecture that enables openness, intelligence, and automated control. 
} The RAN Intelligent Controller (RIC) 
provides the platform 
to design and deploy RAN controllers.  
xApps are the applications which will take this responsibility 
by leveraging machine learning (ML) algorithms and acting in near-real time. 
Despite the opportunities provided by this new architecture, the progress of practical artificial intelligence (AI)-based solutions for network control and automation has been slow. This is mostly because of the lack of an end-to-end solution for designing, deploying, and testing AI-based xApps fully executable in real O-RAN network. 
In this paper we introduce an end-to-end O-RAN design and evaluation procedure and provide a detailed discussion of developing a Reinforcement Learning (RL) based xApp by using two different RL approaches and considering the latest released O-RAN architecture and interfaces. 
\end{abstract}

\IEEEoverridecommandlockouts

\IEEEpeerreviewmaketitle


\section{Introduction}
Telecommunication networks will soon 
provide wireless connectivity and facilitate tiny and huge data transactions among 10s of billions of smart devices. 
Resource, data, and network management is becoming more challenging and artificial intelligence (AI) solutions are being researched for facilitating future wireless network operations. The major obstacles 
are the resource restrictions and the lack of a proficient platform to 
handle AI solutions completely independently from the network hardware, decreasing the cost of changes to new third-party software solutions \cite{9625494}. 
The hardware, software, and interfaces of traditional radio access networks (RANs) are 
tightly coupled. 
Recent advancements of RAN technology can help breaking such closed designs and vendor monopoly \cite{8869705}. A new architecture introduced by the Open-RAN (O-RAN) Alliance can bring this idea into reality and change the future of RAN deployment, operation, and maintenance~\cite{abdalla2021toward}. 

The O-RAN Alliance defines  specifications to facilitate AI integration and allow machines and software to function intelligently in a cellular network. The O-RAN architecture enables intelligence and openness by providing an infrastructure for integrating RANs on open hardware with embedded AI-based software \cite{9627832}. This architecture supports the Third Generation Partnership Project (3GPP) and other industry standards. 3GPP defines the radio and network protocols for the user equipment, RAN, and core network. O-RAN leverages those and the 
Radio Unit (RU), Distributed Unit (DU), and Centralized Unit (CU) that 3GPP defines and specifies particular RAN functional splits and open interfaces facilitating practical disaggregation of functionalities and integration from different vendors. 
O-RAN also introduces new architectural components: the Near-Real Time (RT) RAN Intelligent Controller (RIC), the Non-RT RIC, and additional interfaces which pave the way for inserting intelligent network control and optimization applications called xApps \cite{bworld2}. 

The challenges for developing 
xApps and deploying them on real networks include: finding the most efficient AI models suitable for 
very large real-world networks, adopting the most proficient network parameters, and 
testing the AI models in an environment that accurately represents the behavior of real-world networks.
I light of these concerns, this paper details the development and testing flow of a reinforcement learning (RL) based xApp based on O-RAN architecture. We discuss a O-RAN architectural components, interfaces, and workflow to design an xApp. We investigate and simulate different AI solutions to analyze the performance of various RL methods for designing the xApp.

The rest of this paper is organized as follow: Section II introduces the O-RAN architecture and the main components that are relevant to the AI controller design. Section III presents the related work. Section IV discusses the design procedure and the challenges for developing xApps. Section V analyzes different AI models to select the most efficient solution. Sections VI describes the xApp development and Section VII the deployment and results. Section VIII draws the conclusions.


\section{O-RAN Architecture and Key Components for xApp Development}
\begin{figure}[ht!] 
\centering
\includegraphics[width=3.4in]{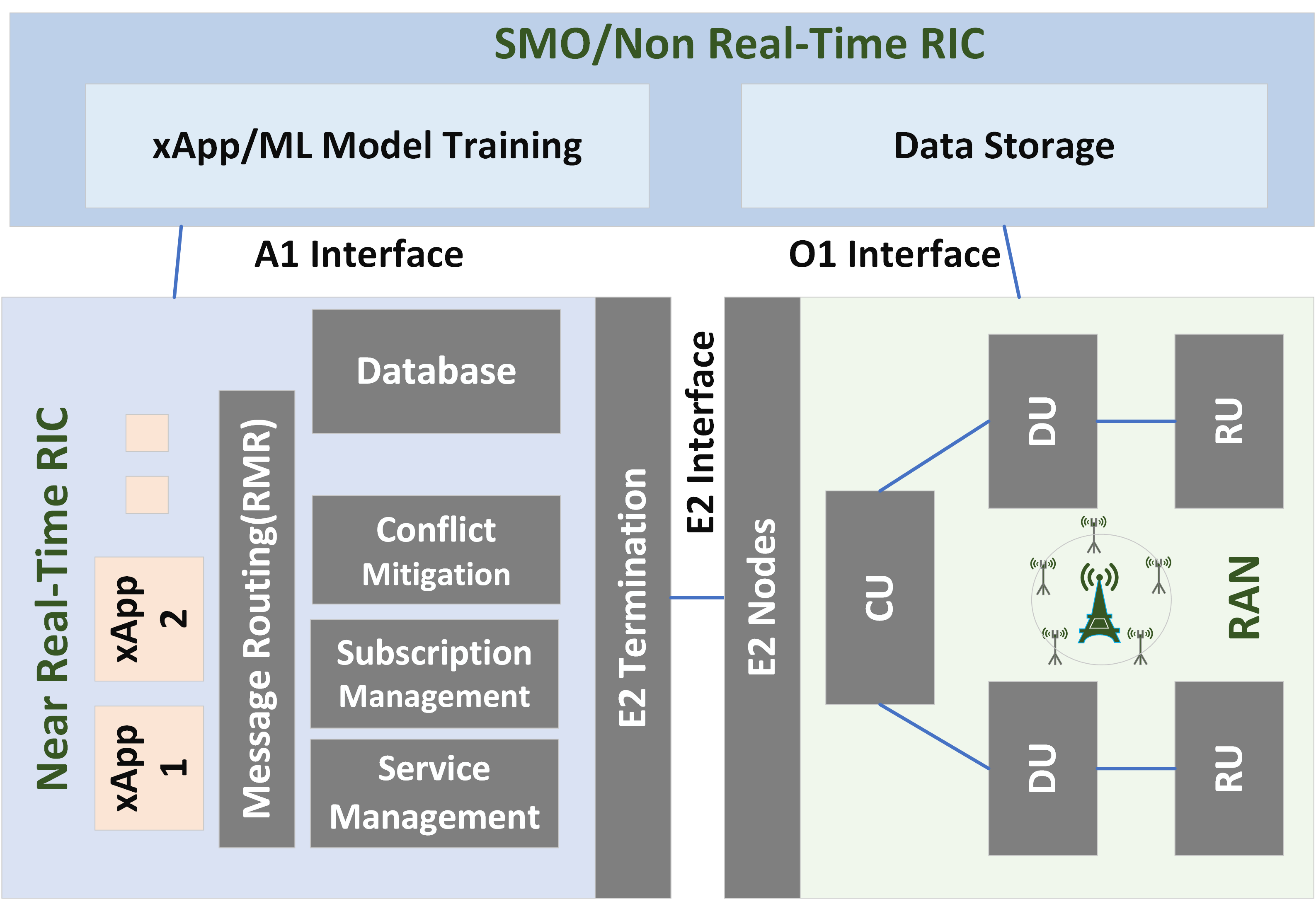}
\caption{O-RAN architecture for RIC application layer.}
\label{RIC}
\end{figure}
The network architecture needs to provide a platform for deploying AI/ML-based applications and 
provide the required infrastructure for data transactions from the RAN nodes to the AI model, data storage, transmission of the model decision and control commands to the network, and the AI model training process. The O-RAN architecture shown in Fig. \ref{RIC} is 
based on 
open interfaces to enable interactions between the RAN and the RAN controller. The RAN is split into three logical units: CU, DU, and RU. The CU is a centralized unit developed to handle the higher layer RAN protocols, such as the radio resource control (RRC), the service data adaptation protocol (SDAP), and the packet data convergence protocol (PDCP). It interfaces with the DUs through the mid-haul. The DU is a logical node that handles the lower protocol layers, which are the radio link control (RLC), the medium access control (MAC), and part of the physical layer (PHY). It interfaces with the RUs through the fronthaul. The RU implements the lower part of the PHY. 

Data is transmitted from the RAN to the non-RT RIC through the O1 interface and is stored in a database for offline training and testing of the AI/ML model. The model training will take place at the non-RT RIC which is also responsible for performing non-RT control operations in O-RAN and for providing and managing higher layer policies. After training, the xApp will run on the near-RT RIC and interact with the RAN through the E2 interface to perform online optimization and control of the network. An xApp can communicate with other parts of the near-RT RIC through internal interfaces which are introduced in Section IV. There is an internal messaging infrastructure called RIC Message Router (RMR) and a shared data layer (SDL) for data sharing. The near-RT RIC provides the framework to handle conflicts, subscriptions, applications, security, and logging.

\section{Related Work}
Several ML based schedulers have been introduced in the literature to address the most challenging problem of resource allocation in cellular networks. Gosal et al. introduce a centralized RL-based scheduler based on the Deep Deterministic Policy Gradient (DDPG) considering pricing rate of resources \cite{9142791}. Elsayed et al. discuss challenges of AI-enabled solutions for optimizing network resource orchestration \cite{8758918}. Polese et al. propose an ML-based edge-controller architecture for the 5G network and use generated data in a testbed to evaluate the model \cite{9107476}. Niknam et al. propose an ML based resource allocation scheme for controlling O-RAN network congestion and evaluate the model using published real-world data from an operator \cite{https://doi.org/10.48550/arxiv.2005.08374}. Mollahasani et al. introduce an optimization method for RL based solutions of resource allocation function in O-RAN \cite{9685837}. They investigate the effects of observation nodes on the performance of the resource allocation. Mollahasani, at al. design a RL based scheduler for allocating resource blocks in a reconfigurable wireless network considering mobility 
\cite{9395230}. Bonati et al. introduce an open experimental toolbox which provides an open testbed for AI/ML xApps and present an ML based scheduler and test results 
\cite{9771908}. 

The prior works focus on the ML application, testing, and evaluation parameters. This paper presents the detailed process for designing and deploying AI/ML based xApps on the RIC. 
The O-RAN Software Community (OSC) has published several xApps developed by its members \cite{bworld4}. 



\section{xApp Development}

To write any AI/ML solution in the format of an xApp to be deployed in 
the RIC, two main steps should be considered. The first is to write an xApp with the essential libraries and functions based on the RIC requirements. For this purpose, developers can use the RIC utility libraries, such as the RMR, SDL, logging, or use a predefined xApp frameworks that have been developed based on the RIC platform requirements. The 
xApp frameworks simplifies developing xApps in Python, go, or C++ 
\cite{bworld4}. We have used the ricxappframe 3.2.0 provided by PyPi for our development to facilitate adding essential features such as communication functions with RMR and SDL. The second step 
is building and deploying the application on Kubernetes since the RIC cluster is developed on the Kubernetes platform. 

\begin{figure}[ht!] 
\centering
\includegraphics[width=3.5in]{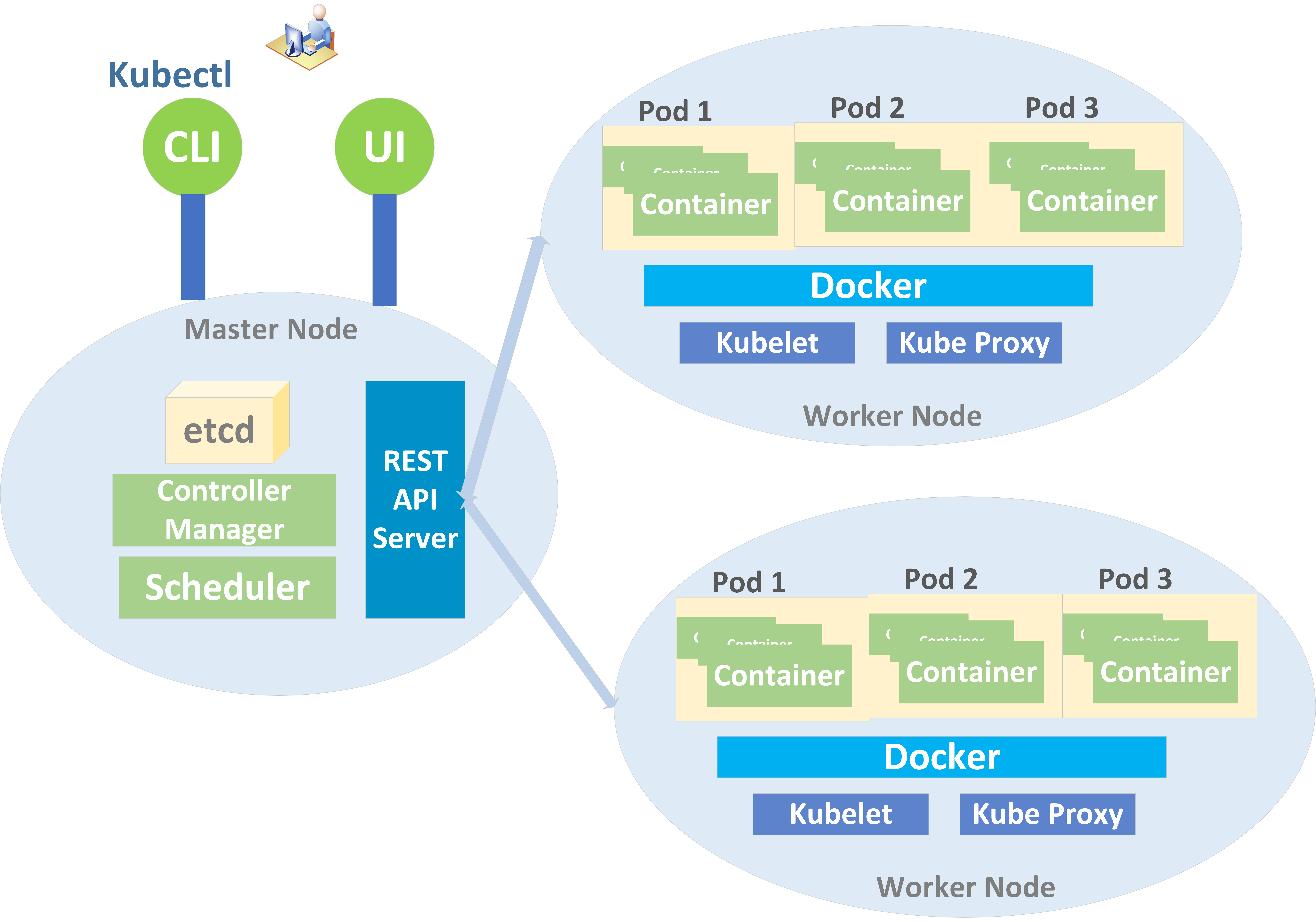}
\caption{Kubernetes Architecture}
\label{kube}
\end{figure}

{Kubernetes} is an open-source platform for deploying and managing containerized applications across clusters of nodes. 
Fig. \ref{kube} shows the high-level Kubernetes architecture which 
consists of a control plane, a master node, and a number of worker nodes that execute the deployed applications. 
The master node hosts the API server, scheduler for assigning worker nodes to applications, etcd as a key-value distributed storage system, 
and a controller manager. 

The worker nodes which are running containerized applications are built of different components. The deployment unit of Kubernetes is Pod, which is a group of containers with shared resources. Pod with all of its containers is deployable through a Yaml file that determine the Pod configurations such as ports, name, the number of replicas and is implemented on one machine that has a single IP address shared among all of its containers. The next component is the container runtime (e.g., Docker) that is responsible for running containers. The Kubeproxy unit routes traffic coming into a node from the service and forwards work requests to the correct containers. 
To provide communication with the master node and containers of the worker node, kubernetes uses kubelet service that also 
traces the states of a pod to check whether all the containers are running.

After developing the main application code in Python within the xApp framework, we need to deploy it as an containerized application on Kubernetes. In RIC To deploy our containerized application we use the ricxapp Pod in the Kubernetes node. For xApp deployment we have four main steps: First we should containerize the application to create a container image. This facilitates porting an application and executing it on any machine. 
To create a container image, we wrote a Docker file that includes the instructions to run our Python code and built our Docker image with that. In the next step we should tag and push the container image to a cloud repository such as Dockerhub. 
Then, In order to deploy the xApp we need to create an xApp descriptor that is a JSON file contains the main configuration parameters required by RIC platform.



\section{proposed xApp architecture}

The next step in designing an xApp is to determine the network nodes, main data-flows, and the architecture. The main concern is designing a deployable platform for data transactions 
between the near-RT RIC and the 
RAN via control messages. Since all the needed data related to the RAN are available on the E2 interface and the xApp can send decision to the RAN through E2 Control Messages, the E2 termination is the main node and E2 messaging procedure will be the main data transaction procedure in the near-RT RIC for our design.

\textbf{E2 Messaging Procedure:}
E2 messages are divided into two classes of support and service types. The interface management functions E2 Setup, E2 Update, and Reset are related to the support class. The RAN CONTROL and REPORT functions are related to the service class. The E2 messages are provided in the Abstract Syntax Notation One (ASN.1) format which is a standard interface description language. Since ASN.1 format is the final encoding format of the messages, any other system that employs the same ASN.1 can easily decode it. This specification has removed the dependency of the system to any specific vendor or language. 

\begin{figure}[ht!] 
\centering
\includegraphics[width=3.5in, height=1.8in]{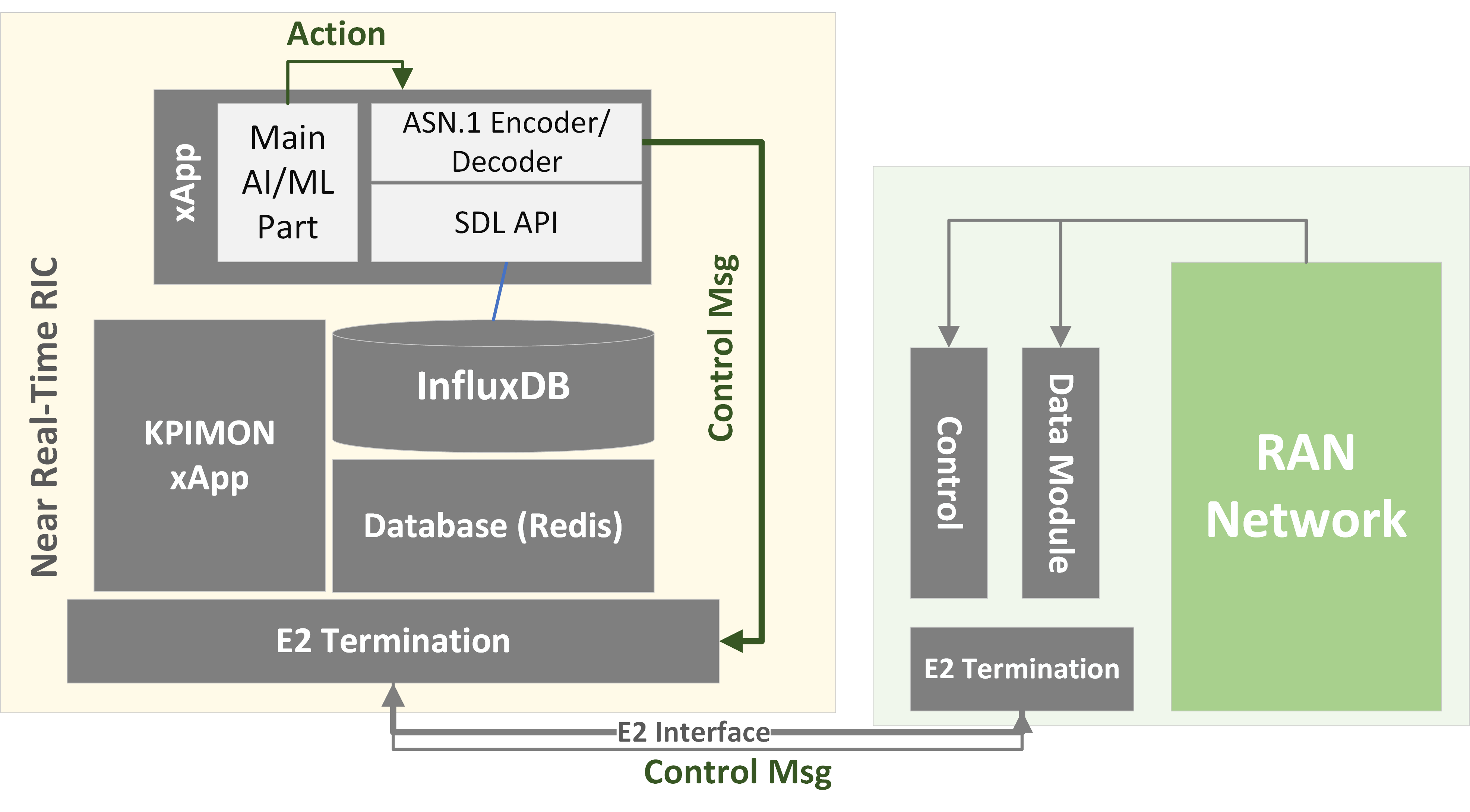}
\caption{xApp architecture with implemented O-RAN modules and interfaces.}
\label{xApp}
\end{figure}

The KPIMON xApp enables collect metrics from E2 nodes. The RAN nodes’ metrics, such as the number of used and available Physical Resource Blocks (PRBs), the number of connected UEs, the downlink and uplink data rates, and so on are collected by the E2 agent. These metrics are packaged in containers. Each container has its own ID with a header to determine the related RAN node (CU, DU, …). These metrics are compiled to form the indication message and will be encoded using ASN.1 encoding. The KPIMON xApp periodically receives these indication messages and uses the same ASN.1 and service model definitions to decode them and get the metrics. KPIMON uses Redis for data storage. Redis is an open-source in-memory data storage that is used as the RIC database. In order to share metrics between KPIMON and the new xApps, a time series database like influxDB can play the role of the sharing layer. Fig. \ref{xApp} shows the designed architecture for the resource allocation xApp.



InfluxDB is a time series (TS) open source database. 
It makes it possible for developers to store, retrieve, and work with TS data that is used for real-time analysis. 
These kind of DBs are especially useful in situations such as monitoring and operations on the logs and metrics of large networks. For our purpose we use the influxdb-client module in Python that is supported by InfluxDB 1.8 and Python 3.6 or later versions.


\section{AI Model Selection and Formulation for Designing xApps}
The next step is to find an AI/ML model that suits the problem. Generally, AI/ML solutions are divided into three classes: Supervised learning, where a set of labeled data is available, unsupervised Learning, where the task is to find a structure of similarity among unlabeled data, and RL, where the learning is based on trial and error while facing with an unknown environment. The best method should be selected based on the problem that the AI/ML model needs to solve. The 
RAN operates in a dynamic environment of variable characteristics, where mobile users with non-stationary channels generate service requests. 
This can lead to different sets of undefined resource allocation actions for which RL would be the appropriate solution.

RL is a learning model designed based on interactions between the agent and the environment. It is often used in control or resource management problems because it can learn from direct interactions with the environment. Each time the RF agent applies an action to the environment, the new state will determine the reward of the system. RL is based on the 
Markov decision process (MDP). It has a state space that defines a series of states $s$ with a distribution 
$p(s)$, a series of actions $a$, state transitions: \[T(S_{t+1}\mid S_t, A_t )\] from time slot $t$ to time slot $t+1$,
 a reward function which is a function of the current state-action and the next state $R(S_t$, $A_t$, $S_{t+1})$, and a discount factor $\gamma$ 
defined between 0 and 1. The policy $\pi$ 
captures the probability distribution of actions and is used by an agent to decide which action is performed in the given state. In other words, 
it is a function that has state $s$ as the input and returns an action $a$ as the output: $\pi$(s) → $a$. In every execution of the policy, the system gathers rewards from the environment. The goal of the RL model is to find an optimum policy $\pi^*$
to earn the maximum reward from the environment across all states:
\begin{equation}
    \begin{split}
        \pi^* &=\max_\pi E\bigg[\sum_t R[_{t+1}|\pi(s_t)] \bigg].  \\
    \end{split}
\label{eq2}
\end{equation}



RL algorithms are categorized in two main classes: value function and policy search. For the value function approaches, the model tries to maximize the value function $V$ or the equivalent $Q$ function by finding some policy $\pi$. Based on the MDP, if the optimal policy is $\pi^*$, the model will act optimally and the best policy in each state can be found by choosing the action 
with the maximum value in each state: 
\begin{equation}
    a^* = \arg \max_a Q_{\pi^*}(s, a).\\
\label{eq3}
\end{equation}





For the direct policy search, 
the model 
searches in the policy space to find the optimal policy $\pi^*$ without the value function model. 
{$\pi$} 
will be chosen by the system as the policy that the RL model 
updates 
to maximize the expected return $E[R|\theta]$.

One concerns with the direct policy search approach is 
the slow convergence when the data is noisy. This usually happens in episodic processes as the case for resource allocation that is a continuous process where the variance can be large \cite{9685837}. 
We can combine the value-function with the policy method to overcome the problem. It is a tradeoff for the variance reduction of policy gradients by introducing bias from the value function. The method is named actor-critic in which the actor network learns the policy using feedback from the critic network that learns the value function. 

As described above the states, actions, and 
return function play important roles in any RL model along with determining the environment accurately. The first step in designing an RL solution is thus to decide about the environment and its characteristics, which is the action and observation place, that must be understood to design a 
high-fidelty model that contains the exact definition of states, actions, reward function, and the flows.

\subsection{Reward Function}

Defining a reward function play a very important role in any RL solution since RL models try to maximize the reward. So, designing an appropriate reward system can establish an efficient RL model to reach the main goals. In our proposed scenario in an environment with multi base Stations (BSs) and multi users, system should decide to connect which User Equipment (UE) to which BS to maximize the global Quality of Experience (QoE). From UE point of view to increase the QoE, any UE intend to get as many resources of available BSs as possible in order to increase its data rate. But, from BS point of view, there is a restricted resource blocks that should be scheduled to different competitive UEs with different network states, which will be discussed in the following, to maximize the network’s QoE. In our QoE-based scheduling model where the UEs are moving, the target of the model is to connect as many UEs as possible ($C$) to the most suitable BSs to maximize the overall data rate and fairness ($fa$). Now, we can write the reward function as:

\vspace{-1mm}
\begin{equation}
    r_{ue} (s,a) = R_{ue}[C, fa]*(\min fa / \max fa)\\
 \label{eq3}
\end{equation}
In each time slots, if model could achieve the target we will return a full reward but for any partially achievements we will assign a degraded reward to the action and for any failure system will feedback a penalty.

\subsection{States}

Since the task of our xApp is scheduling of resources to maximize the network's QoE, the xApp 
needs to decide which Resource Group Block (RGB) to dedicate to which UE from the group of UEs with active resource requests. Here we assume four important network parameters: 
Channel request, 
Channel Quality Indicator (CQIs) from the Channel State Information (CSI), data rate, and UE fairness ($fa$). These are the observation to decide about the resource allocation.

Modulation and Coding Scheme (MCS) defines the number of useful bits that can be transmitted on a Resource Element (RE). It depends on the channel, i.e. CQI, and determines the data rate. 
We use the MCS table from the 5G physical layer specification version 16.2.0 
for mapping.

For UE fairness a matrix will be constructed to trace the history of each UE and update it when new resources are allocated. So, we have a log for all the UEs and their allocated resources. The model tries to keep all UEs in an approximate equality in terms of allocated RGBs in the long term 
for maximum fairness. 

%


\subsection{{Advantage Actor-critic (A2C)}}

The Actor-critic model is a temporal difference (TD) learning algorithm. 
In this approach, the actor network sets the policy that represents a set of possible actions for a given state, and the critic network embodies the estimated value function that evaluates actions. A series of rewards in each time slot $t$ are gathered and multiplied by a discount factor $\gamma\in(0,1)$  to build a series of expected return 
\begin{equation}
    G_t = \sum_t \gamma ^t r_t.\\
 \label{eq3}
\end{equation}

The loss function of the actor-critic model is a combination of losses of both the actor and critic network
:

\vspace{-1mm}
\begin{equation}
    L = L_a + L_c.\\
 \label{eq3}
\end{equation}
By proposing $s_t$ and $a_t$ as the state and action at time slot $t$, $\pi_\theta$ as the policy parameterized by $\theta$, $V_\pi^\theta$ as the value function (critic) parameterized by $\theta$, 
the policy gradient based actor loss can be written as

\begin{table}[ht]
\centering
\caption{PPO algorithm used in our model.}
\label{tab:ppo}
{\begin{tabular}{|p{8.3cm}|}
\hline
\vspace{0.03 in}
\textbf{PPO Algorithm}
\vspace{0.03 in} \\ \hline
\vspace{0.01 in}
1. Initial guess of the policy parameters ($\theta$) and find value function parameters ($\phi$)

\medskip
2.  For iteration $k$ = 0, 1, 2, ... do:

\medskip

\hspace{0.2cm} 1.  Collect set of trajectories $D_k = [\tau_i]$ by running policy $\pi_k = \pi(\theta_k)$

\medskip

\hspace{0.2cm} 2.  Compute reward-to-go $\hat{R_t}$

\medskip

\hspace{0.2cm} 3.  Compute Advantage estimates $\hat{A_1}, \hat{A_2}, ...$

\medskip
\hspace{0.2cm} 4. Update  policy by maximizing PPO objective through stochastic gradient ascent:

\begin{multline*}
       \theta_{k+1} = \arg \max_\theta \frac{1}{D_kT} \sum_\tau \sum_t \min \big[\frac{\pi_{\theta(a_t\mid s_t)}}{\pi_{\theta_k(a_t\mid s_t)}}A^\pi_{\theta_k(s_t,a_t)},\\
       G(\epsilon,A^\pi_{\theta_k(s_t,a_t)})]
\end{multline*}

\medskip
\hspace{0.2cm}{5.  Fit value function by regression on mean squared error and via gradient descent algorithm:}

\medskip
\hspace{0.2cm}
$\phi_{k+1} = \arg \min \frac{1}{D_kT} \sum_\tau \sum_t \big[V_\phi(s_t) - \hat{R_t}]^2$

\medskip

3. end For

\vspace{0.05 in}
\\ \hline
\end{tabular}%
}

\label{tab:xapps}
\vspace{-1mm}
\end{table}

\vspace{-1mm}
\begin{equation}
    L_a = - \sum_t \lg_ {\pi_\theta} (a_t \mid s_t) [G(s_t, a_t) - V_\theta ^\pi(s_t)].\\
 \label{eq3}
\end{equation}
\vspace{3mm}

The term $(G - V)$ is called advantage which indicates how much 
is 
earned in each state by selecting a random action based on policy $\pi$. By assuming critic as the baseline, the system is trained to decrease $(G - V)$, which leads to lower variance.



\subsection{{Proximal Policy Optimization}}

Proximal Policy Optimization (PPO) is a policy gradient algorithm that uses the actor-critic model to train a stochastic policy. We propose the PPO method as the second method to compare the solutions and analyze its performance in our actor-critic based resource allocation model. Similar to the actor-critic model, actor represents the actions and the critic gives provides the estimations of the value function for evaluating the actions and predicting the rewards. The PPO algorithm 
collects a series of trajectories in each episode by sampling from the stochastic policy. Then, the policy and value functions are updated based on the rewards-to-go and advantage estimation. Generally, the policy uses a stochastic gradient ascent as an optimizer to be updated but the value function which uses the gradient descent to be fitted. Table \ref{tab:ppo} shows the main algorithm we propose for employing the PPO method. 














\section{Deployment and Results}

\begin{figure}[ht!] 
\centering
\includegraphics[width=3.45in]{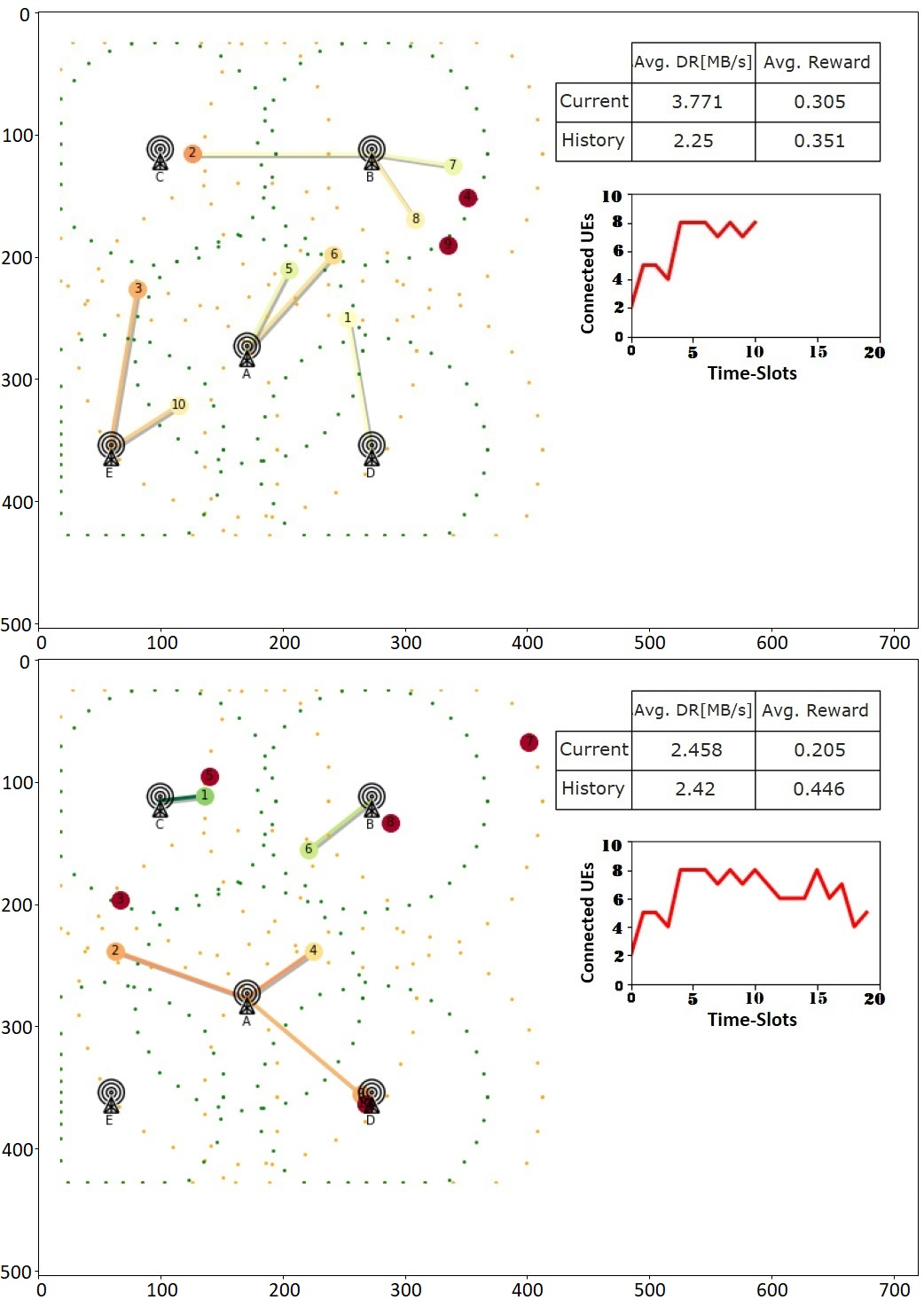}
\caption{Simulated RAN environment, which shows the UE positions, connection states, and data rates in the 10th (top) and 20th (bottom) time slot of the training process.}
\vspace{-1mm}
\label{sim}
\end{figure}

The AI model is developed using Tensorflow2 and expanded to the format of xApp using the ricxappframe framework. We deploy the near-RT RIC as Kubernetes pods with related interfaces provided by OSC to interact with the E2 nodes and the SMO. The proposed 
xApp and KPIMON 
are both deployed in the near-RT RIC and influxDB is used for implementing the shared database.  The implementation is done on Intel i7 11th Gen Processor, with 32 GB of RAM, an Nvidia RTX-3070 Ti GPU to handle AI/ML functionality, 1 TB of storage capacity, and Ubuntu 20.04 operating system. For the training, testing, and evaluation of the xApp, we use the simulated RAN environment designed by Mobile-env \cite{schneider2022mobileenv} that implements an openAI Gym interface. It is written in 
Python and can be installed through PyPi. It provides a 
simulated RAN environment with different customizable scenarios featuring different network configuration such as moving users, multiple base stations, different frequencies, and so on. For our testing scenario, we customize the environment to have 5 BSs and 10 moving UEs. Fig. \ref{sim} shows the five BS towers with their coverage ranges. The movement of the users can be observed from the two snapshots that representing different time slots. The lines between UEs and BSs indicate the connection and the colors 
the user's QoE, where green represents good and red poor QoE.

\begin{table}[ht]
\centering
\caption{Testing Network Parameters.}
\label{tab:Net_Par}
{\begin{tabular}{|p{3.9cm}|p{1.5cm}|}
\hline
\vspace{0.01 in} \hspace{0.02 in}
\textbf{Network Parameters} & \vspace{0.01 in} \hspace{0.06 in} \textbf{Value}
\medskip
\vspace{0.02 in}\\ \hline
Number of Hidden Layers &
\hspace{0.19 in} 4

\\ \hline
Number of Neurons &
\hspace{0.16 in} 384

\\ \hline
Number of gNB &
\hspace{0.19 in} 5

\\ \hline
Number of Ues &
\hspace{0.17 in} 10

\\ \hline
Maximum Traffic load per UE(Downlink)r &
\smallskip \hspace{0.10 in} 1 Mbps

\\ \hline
Frequency  &
\hspace{0.12 in} 3.5 GHz

\\ \hline
Bandwidth &
\hspace{0.10 in} 10 MHz

\\ \hline

Tx &
\hspace{0.10 in} 30 dBm

\\ \hline

Discount factor &
\hspace{0.15 in} 0.9

\\ \hline

Actor learning-rate &
\hspace{0.15 in} 0.01

\\ \hline

Critic learning-rate &
\hspace{0.15 in} 0.04

\\ \hline

Optmizer &
\hspace{0.13 in} Adam

\\ \hline

Number of Timeslots per Episode &
\hspace{0.13 in} 1000

\\ \hline

Number of Episodes &
\hspace{0.10 in} 500-700

\\ \hline
\end{tabular}%
}

\label{tab:xapps}
\end{table}

\begin{figure}[ht!] 
\centering
\includegraphics[width=3.2in]{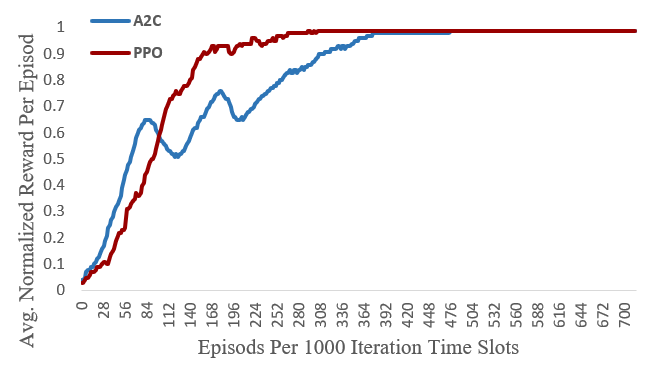}
\caption{Average normalized reward over period for both RL models. 
}
\label{reward}
\end{figure}

\begin{figure}[ht!] 
\centering
\includegraphics[width=3.6in, height=1.1 in]{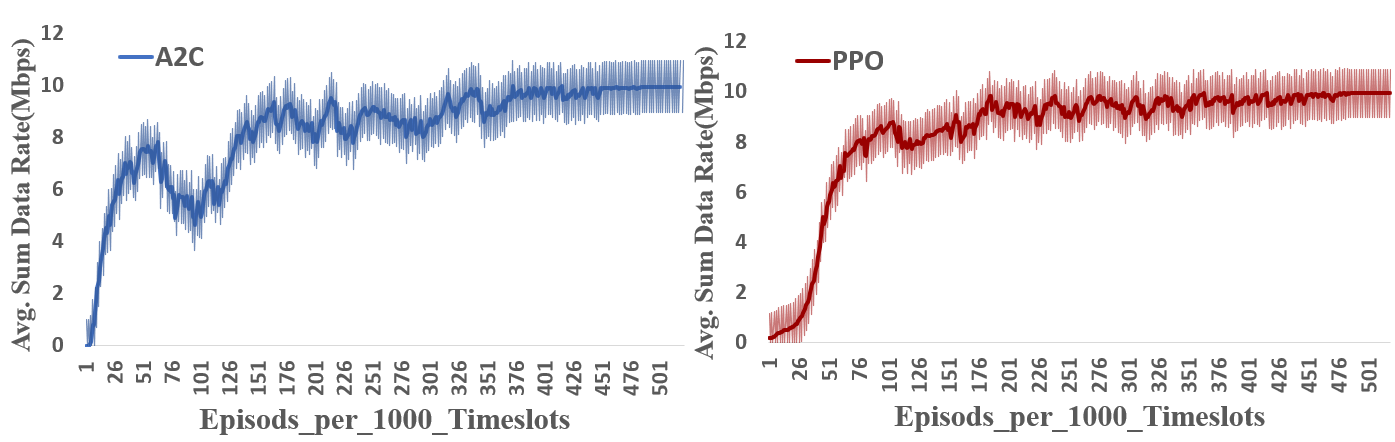}
\caption{Average sum data rate for both RL models. 
}
\label{datarate}
\end{figure}

Table \ref{tab:Net_Par} provide the details of designed A2C network and simulation parameters. The scheduling 
executes every TTI. The results are related to an average of 500 to 700 runs with 1000 iterations.  In order to evaluate the performance of the proposed xApp, since the model is designed to maximize the QoE based on the reward function, we consider the sum of average data rate and rewards as the two metrics for our evaluation. Since reaching convergence in fewer number of iterations is especially important for online training, the required time for convergence should also be considered in selecting the best model. Fig. \ref{reward} and Fig. \ref{datarate} show the results. 

The results of Fig. \ref{reward} and Fig. \ref{datarate} show that both the A2C and PPO models converge successfully. A2C performs better at the early steps 
the PPO model converges quicker and reaches the maximum reward in fewer steps. The results also illustrate the better stability of the PPO model in comparison to A2C. 

We will integrate this xApp development framework into our open AI cellular testbed and share the open-source code and installation instructions through Github~\cite{pratheek2022}.


\section{Conclusions}
In this work we illustrate a step-by-step design, development, and testing of an AI based resource allocation xApp for the near-RT RIC of the O-RAN architecture. The designed xApp leverages RL. The basic version is designed using the A2C algorithm, which is further optimized  using PPO method. The results shows improvements in returned values, metrics, and stability of the system. We plan to extend this work by expanding our test network, and neighboring cell parameters. Also we have plan to handle more tasks through one xApp using new AI/ML methods to increase efficiency while considering task management optimization.

\section*{\textcolor{black}{Acknowledgment}}
\noindent
This work was supported in part by NSF award CNS-2120442.


\bibliographystyle{IEEEtran}
\bibliography{./bib/main.bib}

\end{document}